\documentclass[aps,epsfig,preprint]{revtex4-1}
\usepackage{color}
\usepackage{graphics}
\usepackage{epsfig}
\usepackage{amsmath}
\usepackage{amssymb}
\usepackage[nooneline,font=small]{caption}
\usepackage{subcaption}

\newcommand{\be}{\begin{equation}}
\newcommand{\ee}{\end{equation}}
\newcommand{\bea}{\begin{eqnarray}}
\newcommand{\eea}{\end{eqnarray}}

\begin{document}
\addtocounter{MaxMatrixCols}{7}
\title{Turbulence in two dimensional visco - elastic medium }
\author{Sanat Kumar Tiwari}
\author{Vikram Singh Dharodi}
\author{Amita Das}
\author{Bhavesh G. Patel}
\author{Predhiman Kaw}
\affiliation{Institute for Plasma Research, Bhat , Gandhinagar - 382428, India }
\date{\today}
\begin{abstract} 
The properties of decaying turbulence is studied with the help of a Generalized Hydrodynamic (GHD) fluid model  
in the context of two dimensional  visco - elastic medium such as a strongly coupled dusty plasma system.  
For the incompressible case considered  here however,  the observations are valid for a wider class of visco - elastic systems not necessarily 
associated with plasmas only. Our observations show that an initial spectrum that is confined in 
a limited domain of wave numbers becomes broad,  even when the Reynold's number is much less than the critical value required  
for the onset of turbulence in Newtonian fluids. 
This is a signature of elastic turbulence where Weissenberg's number also plays a role in the onset of  turbulence. 
This  has been reported in several experiments. 
It is also shown that the existence of memory relaxation time parameter and the transverse shear wave inhibit the 
normal process (for 2-D systems) of inverse spectral cascade  in this case.  
A detailed simulation study has been carried out for the understanding of this inhibition. 

\end{abstract}
\pacs{} 
\maketitle 
\section{Introduction}
The understanding of fluid flows has been of great interest in past and a significant amount  of research have been 
dedicated to this particular area  \cite{bachlor_67, gall_fluid}. 
The  interest ranges for studies concerning laminar to turbulent flows, viscous to inviscid 
flows, variety of unsteady flows, fluid instabilities etc. While most of fluid flows have been found 
to follow Newtonian dynamics and hence  standard 
Navier-Stokes (NS) equations applies, a range of fluid  (viz. polymeric and colloidal suspensions, cellular solids) 
have  non - Newtonian dynamics \cite{bird_visco}. Often such fluids  have characteristics  which is a combination of 
 both liquid and solid behavior. For instance in addition to having viscous nature, they 
 also exhibit  solid like elastic features. 
Several models have been proposed for the understanding of  such non - Newtonian fluids. 
The combination of viscous behavior and solid like elastic features (often termed as visco - elastic fluids)  
in some cases have been depicted by introducing a memory relaxation time in the model description. 
Such a model is termed as Generalized Hydrodynamic  fluid model.

The GHD model essentially describes a medium which plays a dual role of the  fluid as well as  also the medium 
having elastic traits due to its strong coupling nature.  It differs in this sense with ``polymeric fluids" 
where the extension field of the polymer additives in a fluid medium provide elastic stresses.  
Such polymeric fluids are often modeled by Oldroyd B kind of models \cite{berti_10} which differ significantly from the GHD model. 
It will, however, be interesting to understand the similarities and differences between  the  GHD model 
studied here as a representative of a visco - elastic behaviour exhibited by strongly coupled systems,  
with that of the Polymeric fluids where elasticity arises through external polymer additives.

 In recent years there have been a great deal of interest in understanding the behavior of such visco - elastic system of  
fluids \cite{goff_13,pan_13,brust_13,habdas_06,lakes_04}. 
The study of the properties of such unique fluids is important 
from the perspective of their interesting transport behavior. The transport properties of any fluid system 
is known to significantly get influenced by the presence of the phenomena of turbulence. It is, therefore, 
both interesting and important to understand the phenomena of turbulence in the context of these 
novel systems.  
Experimental work on the study of 
turbulence have been conducted on  many such visco - elastic systems, which show that the turbulent phenomena 
in the context of such systems are fairly distinct when compared to Newtonian hydrodynamics.  
For instance,  experiments on  Polymeric flows have shown onset of turbulence 
even when the Reynold's number is less than the critical value for the onset of turbulence for normal Newtonian 
fluid systems \cite{Groisman_nature,groisman_njp}. The power spectral behaviour is also different in this case.

As mentioned in the beginning, the visco - elastic medium is often represented  by a Generalized Hydrodynamic (GHD) 
model with a memory relaxation  time parameter to represent elasticity features in the medium. This model 
has been found to adequately represent  various experimental observations in both linear and nonlinear regimes. 
For instance,  the dusty plasma medium in the intermediate range of coupling parameter $\Gamma$ (ratio of inter-particle potential 
energy to its thermal energy)   has both  fluid as well as solid 
like traits.  A  Generalized Hydrodynamic (GHD) description mimicking the visco -elastic 
nature has been employed to study this medium  \cite{Kawghd98}.  
The model has successfully predicted the existence of transverse shear waves which was later 
verified in experiments \cite{Kawghd98,pramanik_2002_prl}.
Numerical evidence for the  existence of transverse shear waves, singular
cusp structures  \cite{sanat_weak_njp,sanat_strong_1d,sanat_kh_str} have also been provided. 
Numerical studies in 2-D have also indicated the formation and the sustained presence of short scale structure, 
a characteristic significantly different than that of the 2-D hydrodynamic flows, which exhibits inverse cascade in two dimension. 

Keeping this in view, the prime objective of the manuscript is to study the phenomena of turbulence in visco - elastic medium 
 by numerically simulating the GHD model description. A detailed study of the spectral cascade behavior in 
 2-D for such systems has  been carried out. We restrict our studies here to the simplified 
 incompressible limit. This helps us in investigating the role of pure transverse shear waves (an attribute of elasticity in 
 the medium) in such studies. 
 The extraneous effects due to compressible fluctuations have been ignored at the moment.  
As the incompressible limit of GHD model assumes no density perturbations, 
in the context of strongly coupled dusty plasma medium this corresponds to the absence of any  charge fluctuations. 
The results obtained here,  
represent a wide range of fluid flows and not merely the strongly coupled plasma system, 
which would be represented with such a  visco-elastic prescription.

The paper has been organized as follows.  A brief description of the governing equations for such an incompressible 
GHD model  and the numerical procedure adopted for 
the simulation studies have  been provided in section \ref{govn}. Section \ref{numob} contains the 
results of our  numerical simulation. 
The theoretical analysis and discussion has been provided in section \ref{diss6}. Section \ref{concl6} contains the conclusion. 
\section{Governing Equations}
\label{govn}
\begin{figure}[!ht]
\centering
\includegraphics[height=10.0cm,width=14.0cm]{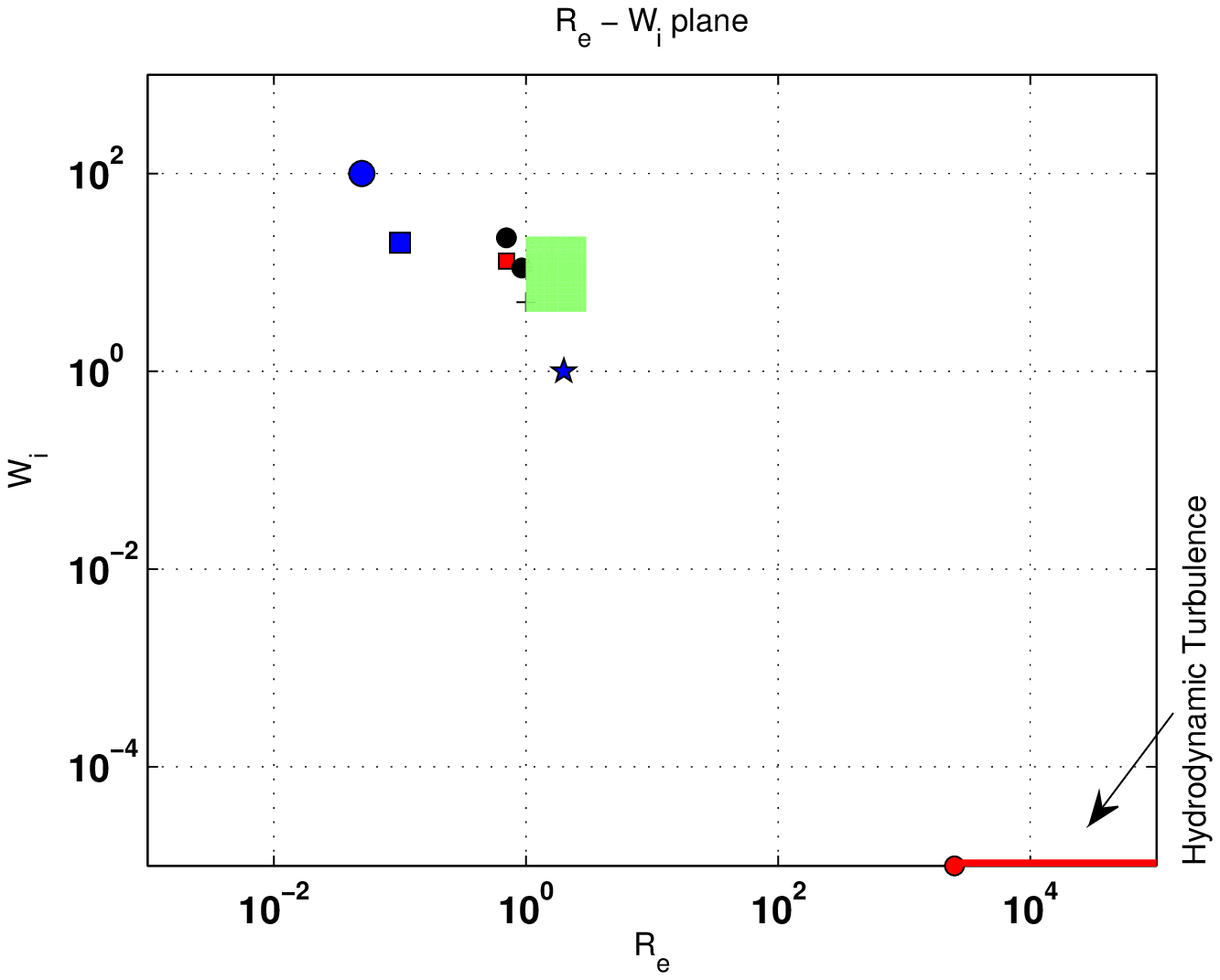}
\caption{ Red line with circle (\textcolor{red}{$\multimap$}) shows the regime of Hydrodynamic turbulence with its 
starting point at $R_e \approx 2500$ \cite{Avila2011}. Black dots ($\bullet$) show the regime of elastic turbulence 
simulation by Berti et. al. using Oldroyd-B model \cite{berti_10}. Blue circle, square and pentagon
(\textcolor{blue}{$\circ,\Box, \bigstar$}) are the points on $R_e-W_i$ plot where authors have contributed 
; the work has been presented here as well as in some previous publications \cite{sanat_kh_str,sanat_kh_aip13}. 
Plus ($+$) is point where Molecular dynamics simulations have been performed
for strongly coupled Yukawa fluids \cite{Ashwin2010}. Red square (\textcolor{red}{$\Box$} point outs regime where some recent experiments
have been performed by Groisman et. al \cite{Groisman_nature}). The shaded square region covers area in which studies made by
Burghelea et. al \cite{burghelea_07}.} 
\label{f6a}
\end{figure}

The behavior of visco - elastic fluids can be represented by a  Generalized Hydrodynamic description. In this description,
the normalized momentum equation is expressed as \cite{frenkel_kinetic,sanat_kh_str}: 
\begin{equation}
\left[1 + \tau_m \left( \frac{\partial} {\partial t}+ \vec{v}  \cdot \nabla \right) \right] \left[ \left(\frac{\partial} 
{\partial t}+ \vec{v}  \cdot \nabla \right) \vec{v}
+ \frac{\nabla  P}{n} \right] = \eta \nabla^2 \vec{v}
\label{ghd_comp}
\end{equation}
In the absence of compressible fluctuations, the equation of state takes the form of 
\begin{equation}
\nabla \cdot \vec{v} = 0
\label{contd}
\end{equation}
Here, $\vec{v}, n$ and $P$ are the velocity, density and pressure fields of the visco - elastic fluid  respectively. 
The parameter $\tau_m$, is known as the time relaxation parameter and is associated with the typical time 
duration for which the visco - elastic medium retains the memory of  past configurations.
Here,  $\eta$ is the  kinematic viscosity, whose finite value causes  viscous dissipation 
and is also responsible for elastic wave propagation in the medium, depending on whether one is 
observing phenomena with time scales longer or shorter than the relaxation time 
parameter respectively.  For a strongly coupled medium the 
parameters $\eta$ and $\tau_m$ are  dependent on each other. The empirical relationship between the two have 
been written down for several strongly coupled fluid systems \cite{Ichimaru1982,Kawghd98}.
For the case of strongly coupled plasmas, the value of these two parameters 
depends on the coupling parameter $\Gamma$ which is the ratio of inter particle potential energy to the average thermal energy of 
the species \cite{Kawghd98}.
The normalization used for the above Eqs.~(\ref{ghd_comp},\ref{contd}) is as follows. The length and time are normalized with 
 $L_0$ and $T_0$, which are respectively the typical length and time scale of the system under consideration.
 The density is normalized with a typical 
density of system $N_0$. These normalizations are conveniently chosen so that one has to primarily 
deal with numbers of the order of unity in simulation and analysis for any system. 
The relaxation parameter $\tau_m$
is also normalized with $T_0$. The normalized kinematic 
viscosity $\eta_N = \eta/mnU_0L_0$ and the normalized pressure $P_N = P/mnU_0^2$, where 
$U_0=L_0/T_0$. We can also see from here that the normalized viscosity $\eta_N = 1/R_e$, is essentially 
the inverse of Reynold's number. 
The Reynold's number being a dimensionless number defined as the ratio of nonlinearity to dissipation, 
characterizes the typical flow of any fluid system. 
Another important dimensionless number associated with   viscoelastic flows is the Weissenberg number $W_i = U_0 \tau_m/L_0 = \tau_{mN}$.
The Weissenberg number is  essentially the ratio of the memory relaxation time to the typical eddy turn over time of the system. 
It thus characterizes the elastic behavior of the medium.
Some recent experiments have shown  that the elastic turbulence can be excited even  in the regime of low $R_e$ 
provided the Weissenberg number $W_i$ is high \cite{Groisman_nature,groisman_njp}. We
have chosen  similar regime of parameter space  for our present studies and observe 
that the spectrum becomes broad asymptotically even if it is initially confined in a limited domain.  
In Fig.~\ref{f6a}, we have made an attempt to locate the values of typical $R_e$ and $W_i$ for 
which turbulence has been observed in the context of a variety of fluid flow systems. 
 We can easily visualize that if we go along the x-axis,
we follow the hydrodynamic flows, these flows for $R_e$  $>10^5$ show turbulent features. 
The lowest possible value of $R_e$ over this axis for which the turbulent flows have been reported is $R_e \approx 2500$
by Avila et. al \cite{Avila2011}. In the $W_i - R_e$ plane, we have indicated points where elastic turbulence studies have been conducted.
In the figure symbols with red filled box show the points where turbulence has been observed in experiments by Groismann et al. \cite{Groisman_nature}
and those experiments conducted by Burghela et al. \cite{burghelea_07}
 square shaded region respectively. Simulations for visco-elastic fluids have also 
been performed using different models and their corresponding positions have been shown by black dots (Berti et al. using Oldroyd-B model) 
and those represented by blue circle, box and star symbols are by Sanat et al. using Generalized Hydrodynamic model 
\cite{berti_10,sanat_kh_str,sanat_kh_aip13}. It is clear from the figure that elastic turbulence occurs for those values of Reynolds number which 
are much smaller than the critical value of Reynolds number necessary for turbulence in pure hydrodynamic systems. 

For the purpose of numerical simulation we reduce the second order Eq.~(\ref{ghd_comp}) in time to two coupled first 
order equations satisfying the following convective forms.
\begin{eqnarray}
& & \left[ 1+ \tau_m \left( \frac{\partial}{\partial t} + \vec{v} \cdot \nabla  \right)\right] \vec{\Psi} = \eta \nabla^2 \vec{v} 
\label{eq-1} \\
& & \left[ \left( \frac{\partial}{\partial t} + \vec{v} \cdot \nabla  \right)\vec{v} + \frac{\nabla p}{n}   \right] = \vec{\Psi}
\label{eq-2}
\end{eqnarray}
In the 2-D incompressible case, with the flow confined in  the 2-D plane normal to the symmetry axis we can represent the velocity field by 
a scalar potential $\Phi$, satisfying $\hat{z} \times \nabla \Phi = \vec{v}$. Here $\hat{z}$ is directed along the symmetry axis. 
Taking the curl of Eq.~(\ref{eq-2}) we have 
\begin{equation}
\left[ \frac{\partial}{\partial t} + \hat{z} \times \nabla \Phi \cdot \nabla \right] \nabla^2 \Phi = (\nabla \times \Psi)_z = 
 \left(\frac{\partial \Psi_y}{\partial x} - \frac{\partial \Psi_x}{\partial y} \right) 
\label{eq-vort}
\end{equation}
We can construct the following evolution equations for the  square integral quantities by taking the scalar product of $\vec{v}$ 
with Eq.~(\ref{eq-2}) and $\vec{\Psi}$ with Eq.~(\ref{eq-1}) respectively
\begin{eqnarray} 
& &\frac{1}{2} \frac{\partial}{\partial t} \int \int v^2 dx dy = \int \int \vec{v} \cdot \vec{\Psi} dx dy  \\
& &\frac{1}{2} \frac{\partial}{\partial t} \int \int \Psi^2 dx dy = - \frac{1}{\tau_m} \int \int \Psi^2 dx dy + 
\frac{\eta}{\tau_m} \int \int \vec{\Psi} \cdot \nabla^2 \vec{v} dx dy 
\label{sq-int}
\end{eqnarray}
The evolution of these quantities are tracked during the course of numerical simulation for the purpose 
of determining the accuracy of the simulation. 
\begin{figure}[ht]
\includegraphics[height=10.0cm,width=14.0cm]{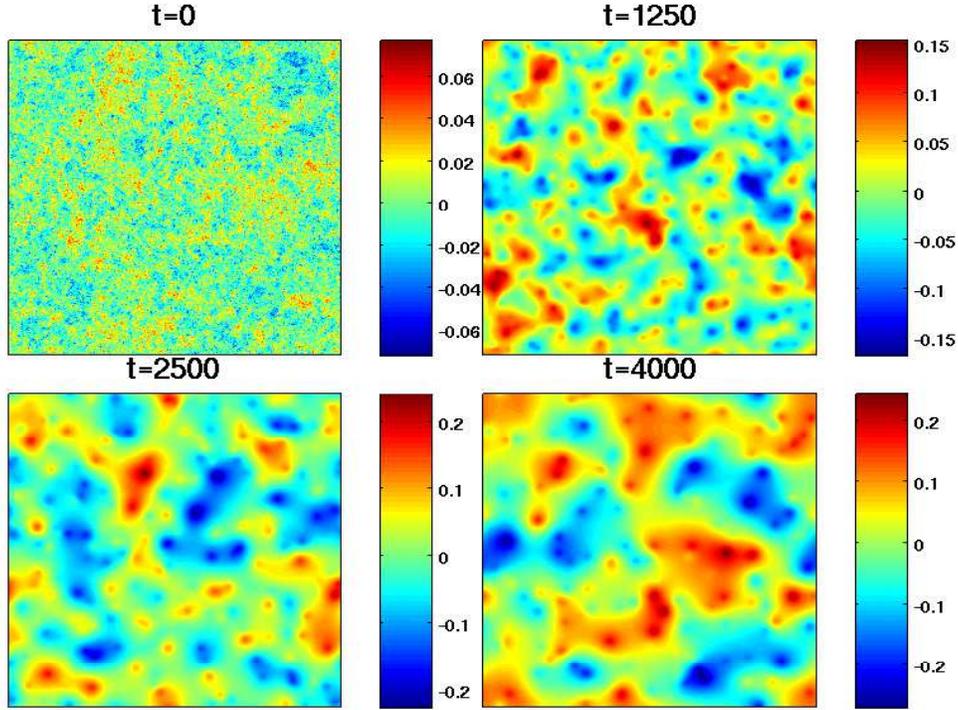}
\caption{Evolution of a random initial configuration of velocity potential $\Phi$ with   the GH code reproducing the incompressible 
hydrodynamic fluid limit for very small values of $\tau_m$ and $\eta$. 
The inverse spectral cascade is clearly evident from the formation of long scale structures.} 
\label{f1}
\end{figure}

A random spectrum of initial fluctuations is prescribed, whose evolution is tracked with the help of numerical simulation. 
For this purpose we employ the following two choices of  initial 
distribution of power in the velocity potential  field $\Phi$ for our studies:
\begin{equation}
\mid \Phi_k \mid^2 = \frac{C}{\left( 1 + \mid k \mid \right)^n}
\label{pw_evol}
\end{equation}
\begin{equation}
\mid \Phi_k \mid^2 = Csech^2\left(\frac{\mid k \mid - k_m}{4}\right)
\label{pw_evol1}
\end{equation}
Here, $C$ is constant which decides the level of power  injected in system. 
The  individual  Fourier modes in the spectrum are chosen to have random phases with spectral weights provided by either by Eq.~(\ref{pw_evol}) 
or Eq.~(\ref{pw_evol1}). 
In the first choice, the power falls of monotonically with wave number (the rapidity of fall being governed by the index $n$)
 and in the latter choice the peak of the initial spectrum can be chosen at any desired 
wavenumber determined by the value of $k_m$. 
Furthermore, in the simulations presented here, we have restricted to an initial choice of 
$\vec{\Psi} = 0$.  These two different choices of the initial spectrum 
help us in ascertaining the dependence if any on initial conditions on evolution. 
This is important in the context of GHD where the medium is expected to retain memory 
for some time. 
These initial conditions are then evolved using the Eqs.~(\ref{eq-1},\ref{eq-vort}) 
through a flux corrected scheme \cite{lcpfct}. 
In the next section we present the details of some interesting observations obtained from our simulation studies. 

We also note that the  general dispersion relation for transverse shear waves obtained from Eq.~(\ref{ghd_comp}) is
\begin{equation}
\omega = \frac{-i}{2\tau_m} \pm 0.5 \sqrt{\frac{-1}{\tau_m^2} + \frac{4 \eta k^2}{\tau_m}}   
 \label{trans_disp_gen}
\end{equation}
This dispersion relation suggests that the transverse waves can propagate in the visco-elastic medium only for those wavenumbers which satisfy the
criteria of $ k > \sqrt{1/4 \eta \tau_m}$. For most of the simulations performed in this paper, we have taken $\eta = 5$ and $\tau_m = 20$ which gives a critical
wavevector $k = k_c = 2.5 \times 10^{-3}$. In our simulations, we have taken $L_x = L_y = 20$ (system length in X and Y direction respectively),
and $Nx = Ny = 1024$ (or $500$ in some cases). Where $Nx$ and $N_y$ are the grid points in X and Y direction respectively. The longest mode 
corresponding to this system will be $k_0 =  0.3142 > k_c$. Thus, in our system, 
the permissible wave numbers are such that the transverse shear mode corresponding to them have propagating characteristic.
\section{Numerical Observations}
\label{numob}
We use the numerical procedure outlined in Section \ref{govn} to reproduce the known case of the hydrodynamic (HD) fluid system. 
For this purpose we have carried out simulations for $\eta = 0$ and chosen $\tau_m$ to be very small. 
In Fig.~\ref{f1} the velocity potential contours for this case has been shown at various times. 
The initial spectrum was chosen to be  provided by Eq.~(\ref{pw_evol}) with n = 2. 
The formation of longer scales at later times are an evidence of  inverse spectral cascade as expected for the 2-D hydrodynamic case. 
A broad  power spectra with indices of -5/3 and -3 in inverse and forward cascade regime has also been recovered for these runs for 
$R_e >> R_e^{crit}$.

\begin{figure}[!ht]
\centering
\includegraphics[height=10.0cm,width=14.0cm]{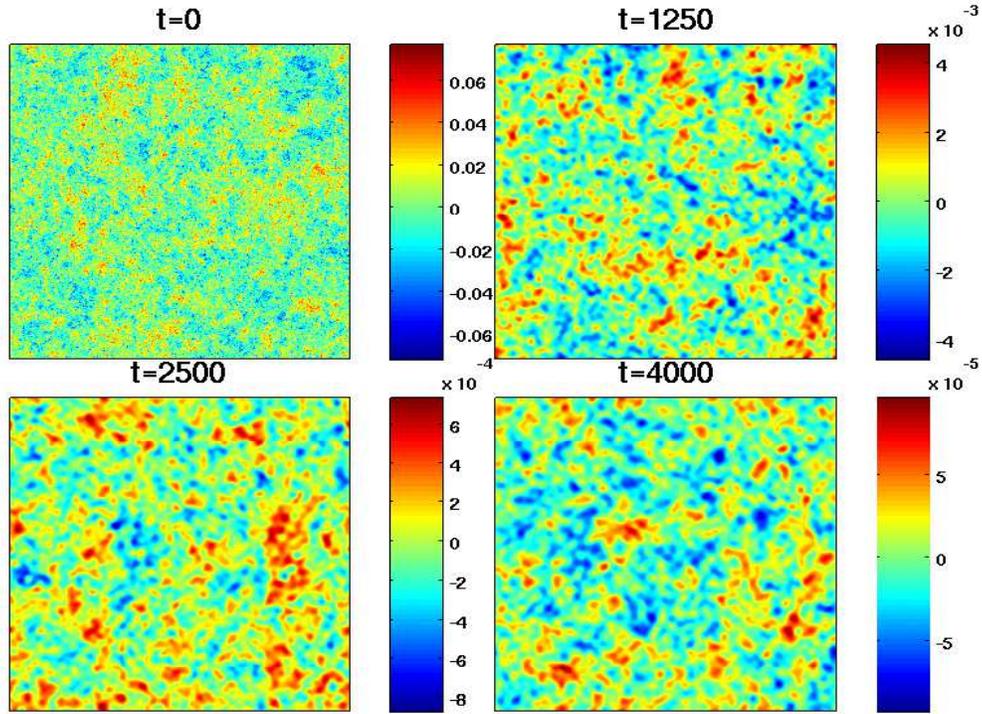}
\caption{Evolution of phase randomized velocity potential for incompressible visco-elastic fluids. It could be seen that in contrast to 
hydrodynamic fluids,  much shorter structures are present in this particular case. 
The parameters $\eta$ and $\tau_m$ are 5 and 20 respectively.} 
\label{f2}
\end{figure}

\begin{figure}[!ht]
\centering
\includegraphics[height=8.0cm,width=12.0cm]{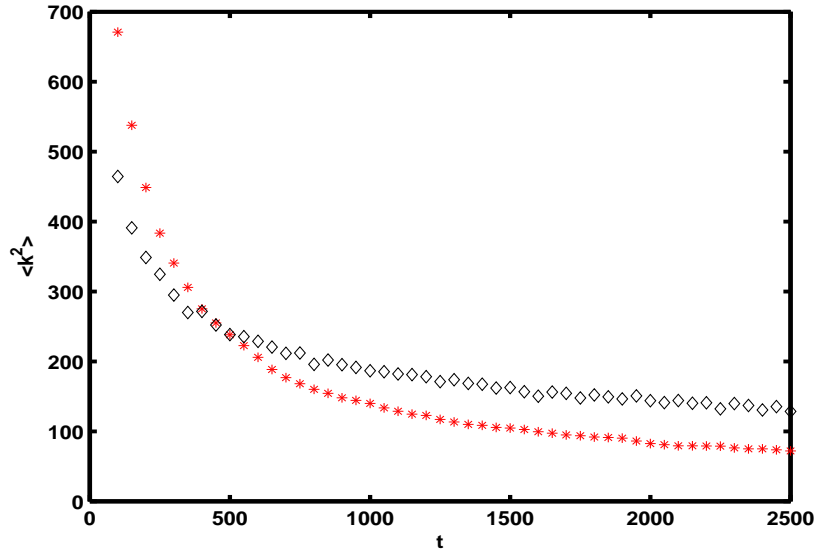}
\caption{The plot of $<k>^2$ with time for the GHD case with ($\eta=5$ and $\tau_m=20$) (square) and for the incompressible fluid (circle).
the initial random flow has been taken of the form given by Eq.~(\ref{pw_evol}) with $n=2$. } 
\label{f3}
\end{figure}

In  Fig.~\ref{f2}, we then show the  constant contour plots of  the velocity potential $\Phi$ for the visco - elastic 
case having $\eta =5,  \tau_m = 20$. 
In this case too, the initial form of velocity potential is given by Eq.~(\ref{pw_evol}) with $n=2$.
It is clear from the figures  that the evolution for the visco -elastic case is   very different. 
For the HD case one  observes a progressive formation of longer  scales which ultimately yielded  
few  isolated vortex structures of the order of system size. 
On the other hand, for visco - elastic case shown in Fig.~\ref{f2}  the time asymptotic state shows a clutter of closely 
packed structures which typically continue to be sustained even though their size is much smaller than that 
of the simulation box dimension. To have a quantitative assessment we have 
 also monitored the evolution of mean square averaged wavenumber $<k^2>$ for the 
 two  cases, which has been shown in Fig.~\ref{f3}. 
It is clear from the figure that  in both  HD and the visco - elastic case described by  GHD cases the spectral cascade is 
towards longer scales as the value of $<k^2>$ is found to decrease. 
However, for the case of visco -elastic system in GHD (now onwards referred as GHD case) this happens much slowly. 
Furthermore, the asymptotic value of the mean square averaged wave number 
is higher (implying sustainment of shorter scales) in GHD as compared  to the HD case. 
Thus, even GHD system seems to  exhibit the 2-D inverse cascade behavior to some extent albeit a bit slowly. It is 
  well known though that in the context of  2-D NS dynamics representing HD flows 
  that the phenomena of inverse spectral cascade is related to  the 
   preservation of  two integral square invariants for by the equations. The GHD equations,however, do not support two integral square invariants. 
~~~~~~~~~~~~~~~~~~~~~~~~~~~~~~~~~~~~~~~~~~~~~~~~~~~~~~~~~~~~~~~~~~~~~~~~~~~~~~~~~~~~~~~~~~~~~~~~~

\begin{figure}[!ht]
\centering
\includegraphics[height=8.0cm,width=10.0cm]{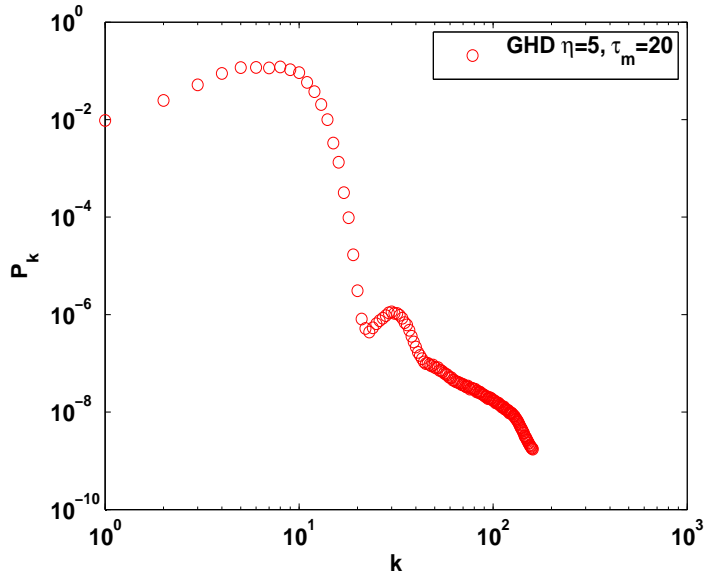}
\caption{Power spectra for incompressible 
visco-elastic fluids ($\eta= 5$ and $\tau_m = 20$).} 
\label{f4}
\end{figure}

\begin{figure}[H]
\centering
       \begin{subfigure}[ht]{1.0\textwidth}
               \centering
               \begin{minipage}[c]{0.54\linewidth}
               \includegraphics[width=\linewidth]{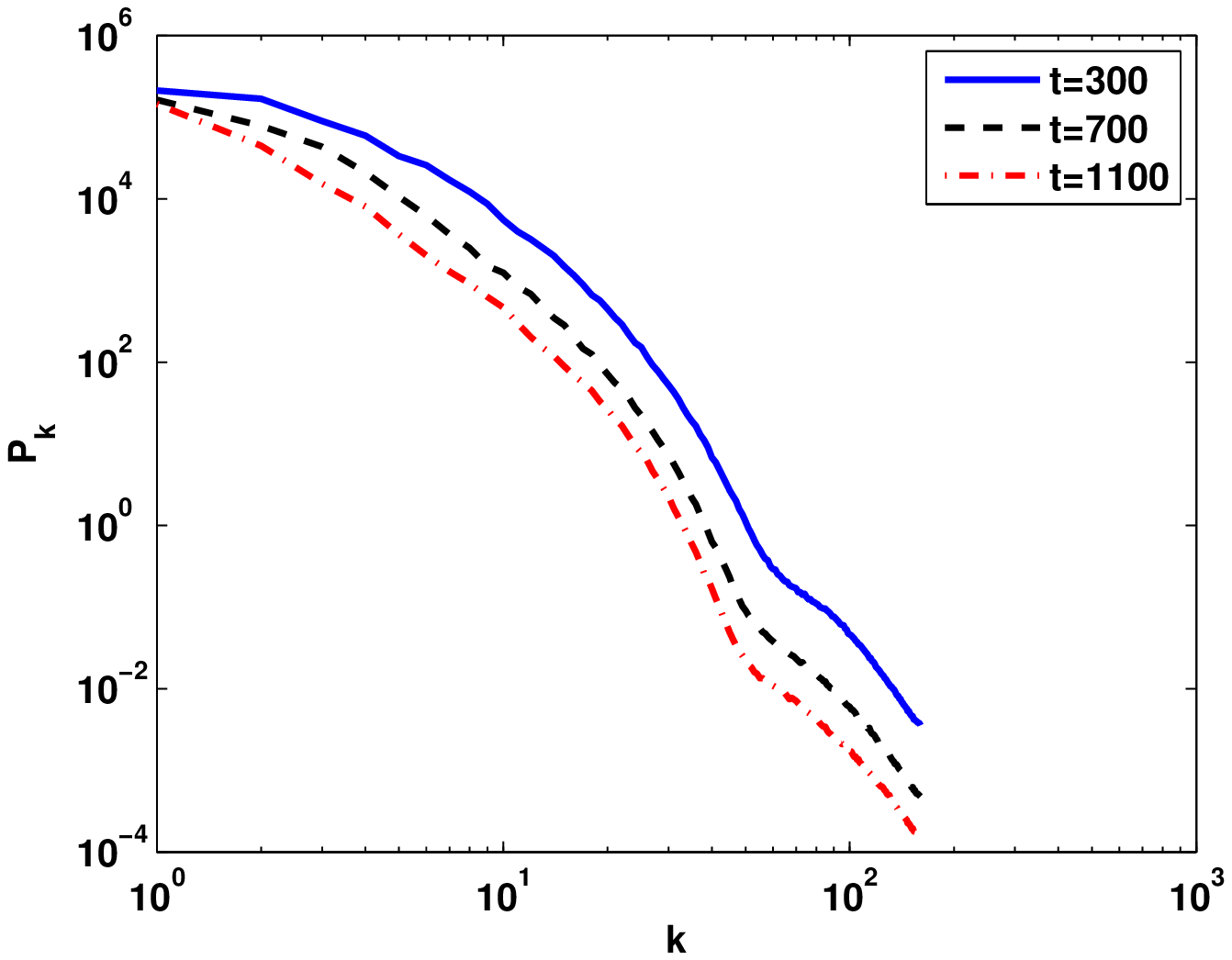} 
               \end{minipage} \hfill
               \begin{minipage}[c]{0.38\linewidth}
                \caption{\large $\mid \phi_k \mid^2 = C sech^2(\frac{\mid k \mid - 20}{4})$ }
               \end{minipage}
               \label{f4b_1}
       \end{subfigure}%
\\
       \begin{subfigure}[ht]{1\textwidth}
               \centering
               \begin{minipage}[c]{0.54\linewidth}
               \includegraphics[width=\textwidth]{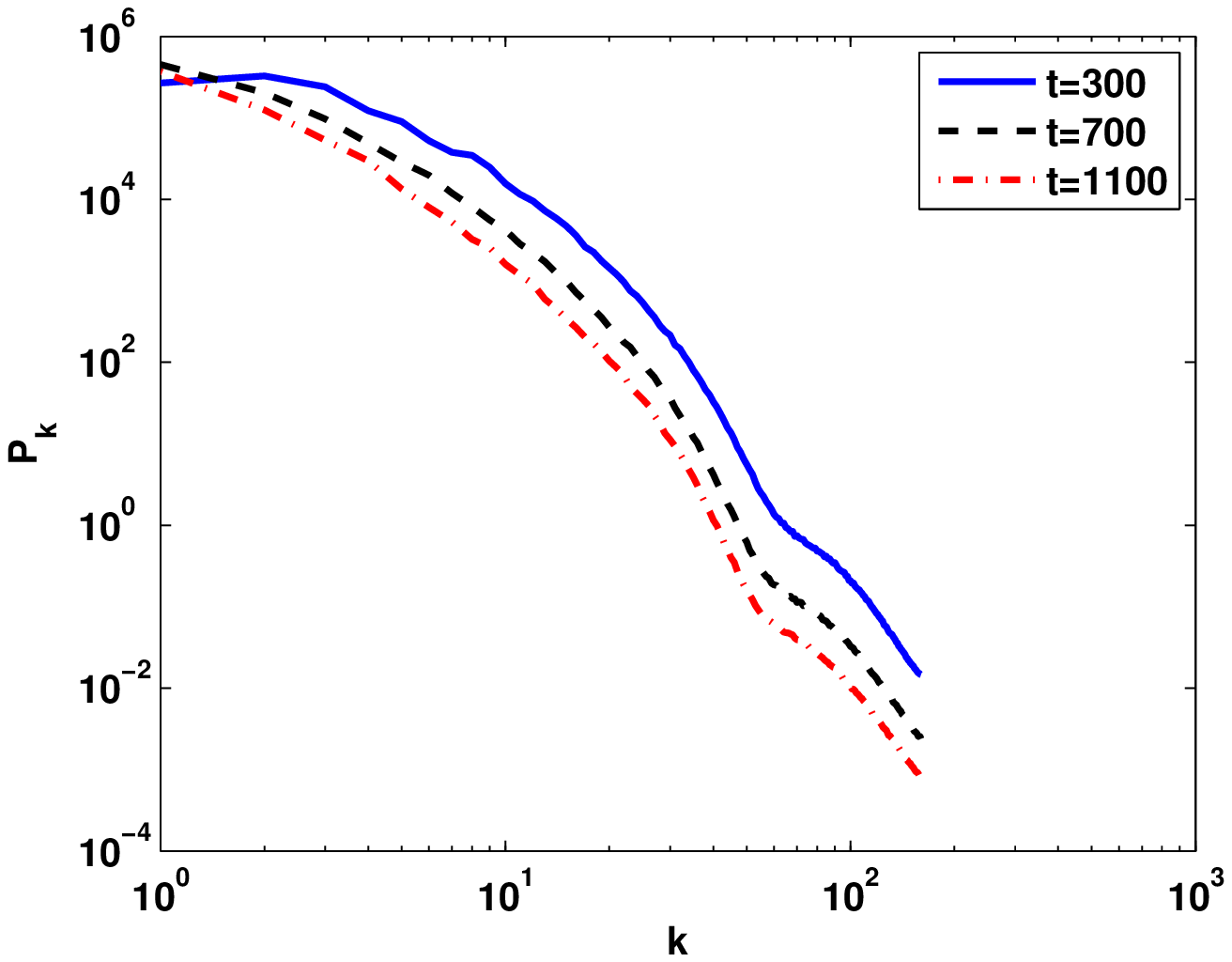} 
               \end{minipage} \hfill
               \begin{minipage}[c]{0.38\linewidth}
               \caption{\large $\mid \phi_k \mid^2 = C sech^2(\frac{\mid k \mid - 30}{4})$}
               \end{minipage}
               \label{f4b_2}
       \end{subfigure}%
\\
       \begin{subfigure}[ht]{1\textwidth}
               \centering
               \begin{minipage}[c]{0.54\linewidth}
               \includegraphics[width=\textwidth]{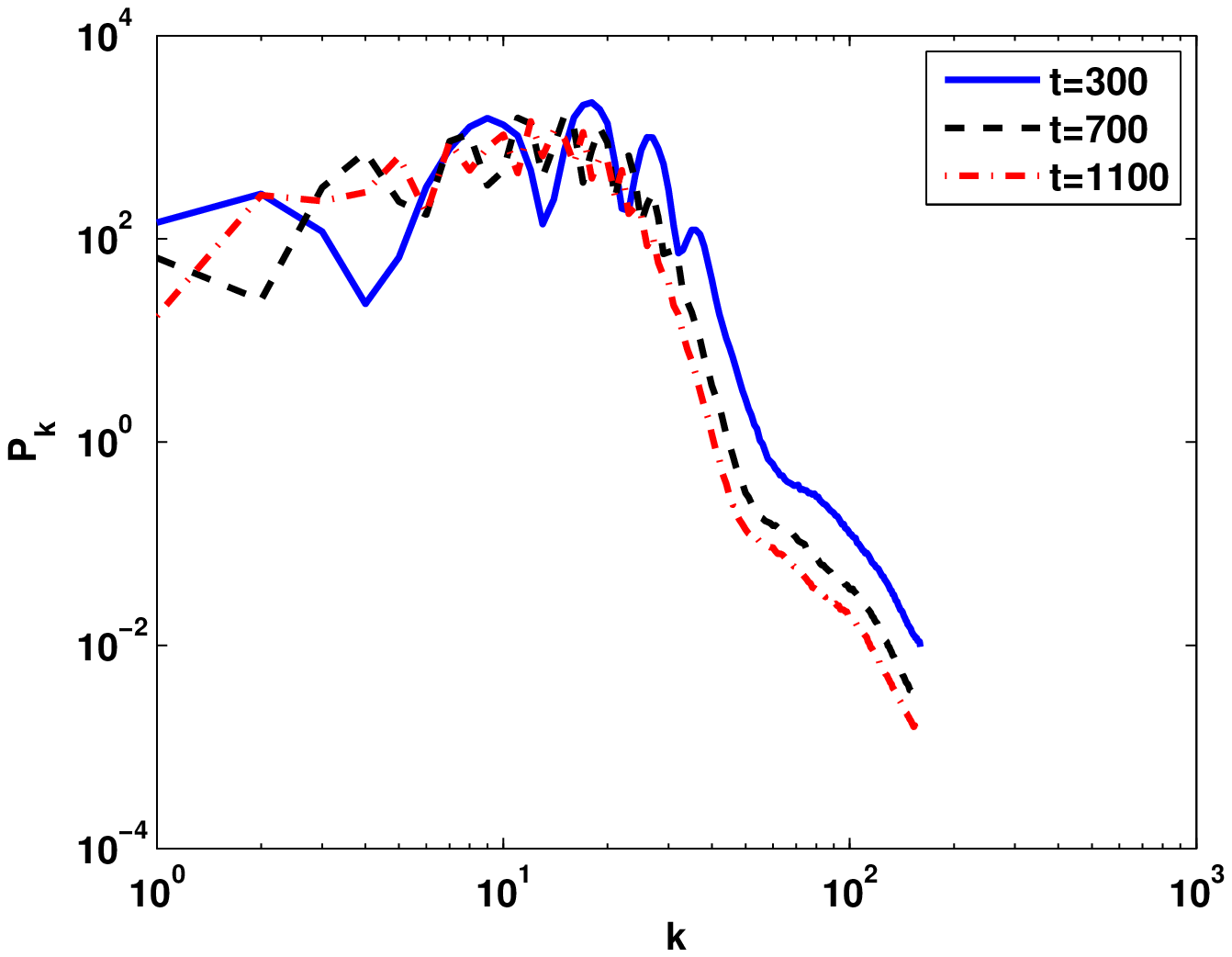} 
               \end{minipage} \hfill
               \begin{minipage}[c]{0.38\linewidth}
               \caption{\large $\mid \phi_k \mid^2 = C sech^2(\frac{\mid k \mid - 90}{4})$}
               \end{minipage}
               \label{f4b_3}
       \end{subfigure}%
\caption{Power spectra for incompressible visco-elastic fluids with different initial form of spectrum mentioned alongside.
For the subplots (a), (b) and (c), the initial power has been injected at $k_m = 20, 30$ and $k_m=90$ respectively.} 
\label{f4b}
\end{figure}

\begin{figure}[!hb]
\centering
       \begin{subfigure}[ht]{1.0\textwidth}
               \centering
               \begin{minipage}[c]{0.54\linewidth}
               \includegraphics[width=\linewidth]{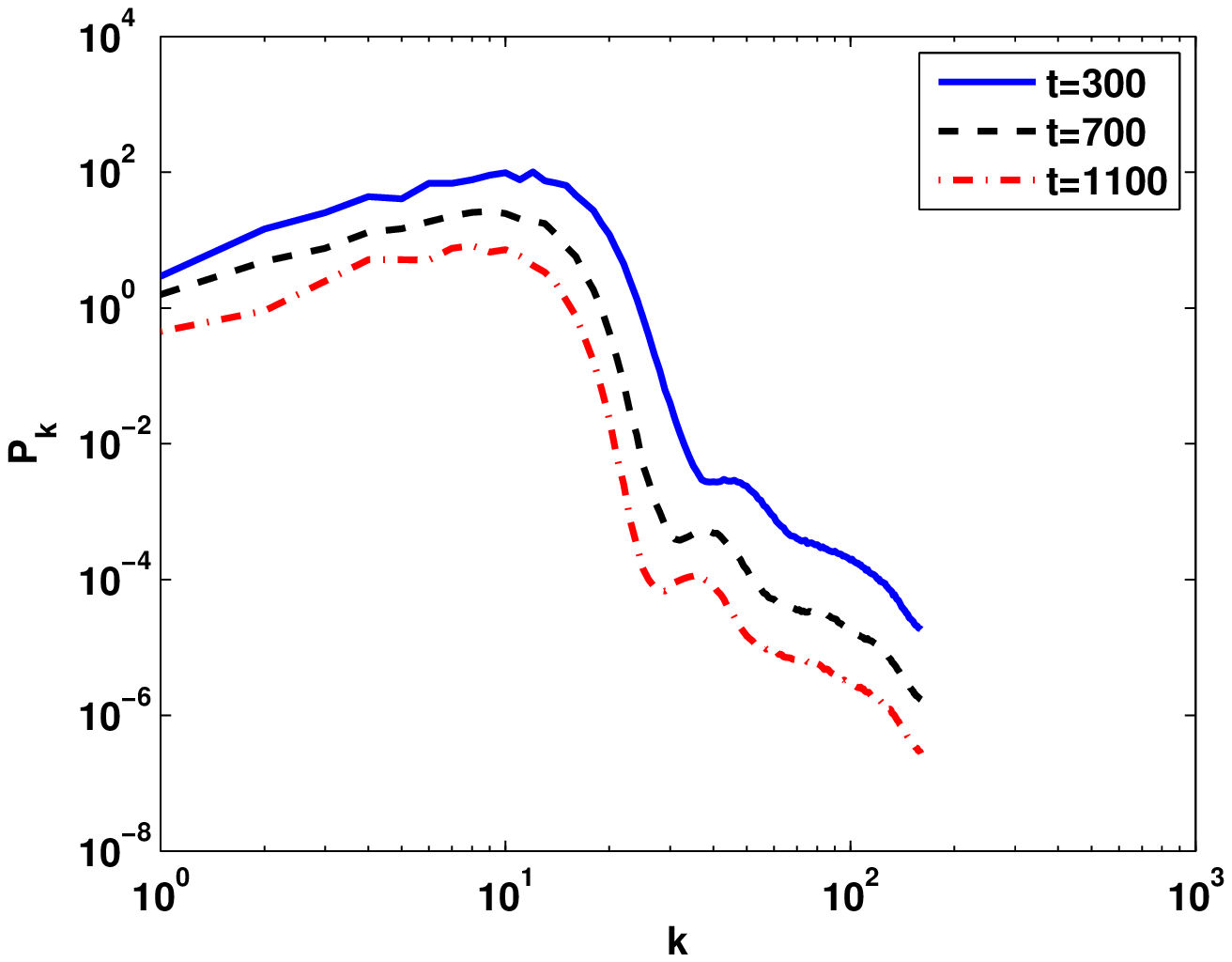} 
               \end{minipage} \hfill
               \begin{minipage}[c]{0.38\linewidth}
                \caption{\large $\mid \phi_k \mid^2 = \frac{C}{\left( 1 + \mid k \mid \right)^8}$}
               \end{minipage}
               \label{f4c_1}
       \end{subfigure}%
\\
       \begin{subfigure}[ht]{1.0\textwidth}
               \centering
               \begin{minipage}[c]{0.54\linewidth}
               \includegraphics[width=\textwidth]{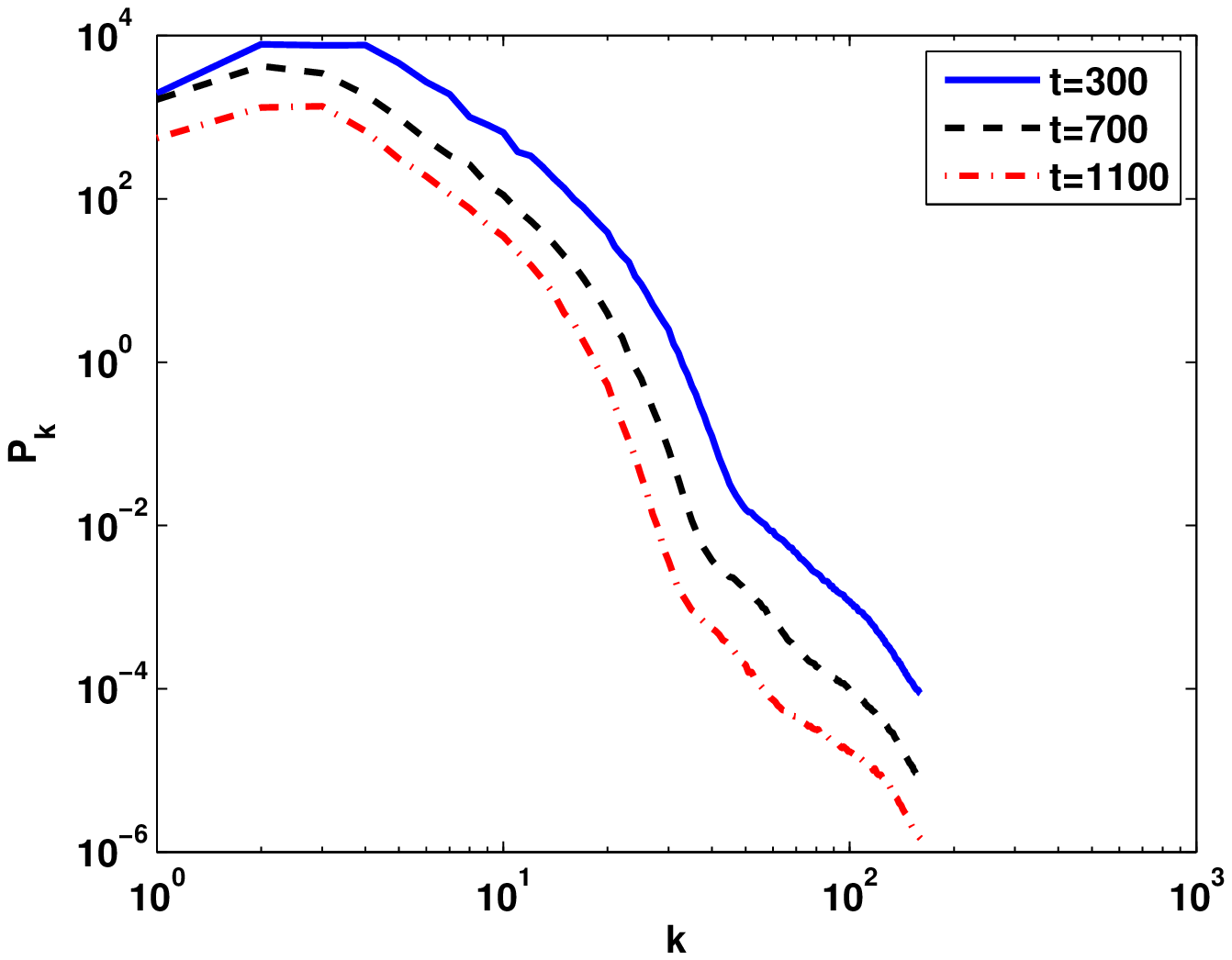} 
               \end{minipage} \hfill
               \begin{minipage}[c]{0.38\linewidth}
               \caption{\large $\mid \phi_k \mid^2 = \frac{C}{\left( 1 + \mid k \mid \right)^4}$}
               \end{minipage}
               \label{f4c_2}
       \end{subfigure}%
\\
       \begin{subfigure}[ht]{1.0\textwidth}
               \centering
               \begin{minipage}[c]{0.54\linewidth}
               \includegraphics[width=\textwidth]{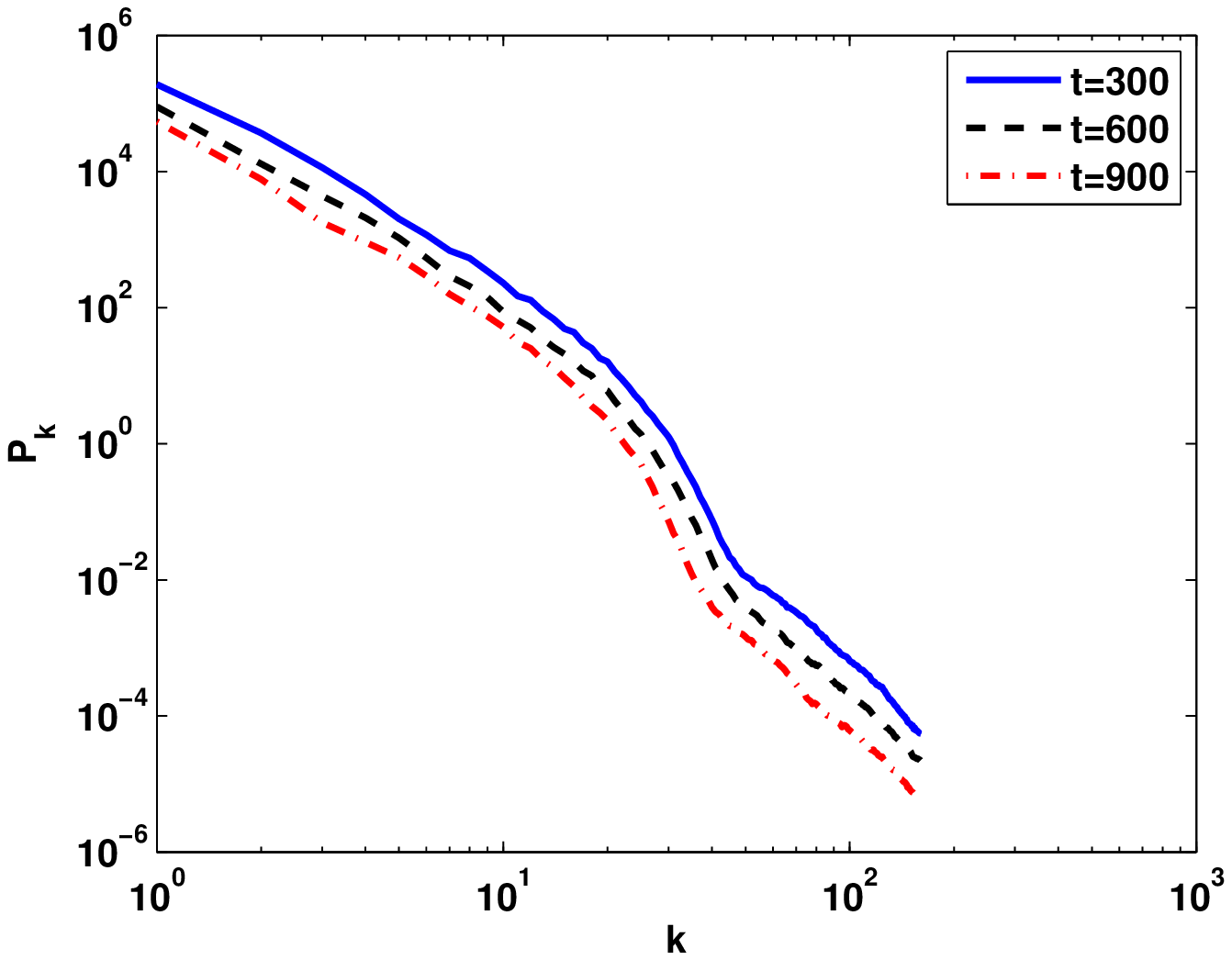} 
               \end{minipage} \hfill
               \begin{minipage}[c]{0.38\linewidth}
               \caption{\large $\mid \phi_k \mid^2 = \frac{C}{\left( 1 + \mid k \mid \right)^2}$}
               \end{minipage}
               \label{f4c_3}
       \end{subfigure}%
\caption{Power spectra with different initial form of spectrum (monotonically distributed power) mentioned alongside.
For the subplots (a), (b) and (c), the initial power has been injected as given in Eq.~(\ref{pw_evol})
with $n=8,4$ and $n=2$ respectively.} 
\label{f4c}
\end{figure}

These simulations being that of  undriven and dissipative nature 
(e.g. in both HD and GHD cases  $\eta$ is responsible for viscous damping, in GHD it 
also has an additional role  of elasticity in the medium ), the overall power spectrum decreases with time. 
However, we observe that with time the spectrum attains an asymptotic form which has been shown in Fig.~\ref{f4}.  
This GHD spectrum is in stark contrast to the HD case. The observed spectrum is broad, 
even though the Reynold's number ($R_e = \eta^{-1} = 0.2$) corresponding to these simulations  is lower than $R_e^{crit}$. 
Here, the Weisenberg number $Wi = 20$. This is a feature of characteristic elastic turbulence and has been reported in 
experimental observations also \cite{groisman_njp}.
The observed broad spectrum, however, does not have any specific  power law dependence. 
The spectrum, instead is  overwhelmed by the  
appearance of peaks  at certain wave numbers. This indicates that the cascade process is severely inhibited 
at certain  wave numbers, resulting in the piling up of the spectral  power near certain wavenumbers location. 
The peaks are not sharp though and their location and shape is found to depend   on the initial choice of the spectrum. 
This is evident from a detailed study carried out with different  choices of the initial spectrum, e.g. using the two forms 
shown in Eqs.~(\ref{pw_evol}) and (\ref{pw_evol1}) with different values of parameters $n$ and $k_m$ chosen for the study. It is clear from 
the  Figs.~\ref{f4b} and  \ref{f4c}
 that for each and every   GHD case, there  is a perceptible 
 shift of the  location of the peak towards longer scales with time.
 It is also clear from the plots that the evolved spectra  depends on the initial choice of 
 spectral power distribution.  We observe that the peak representing the accumulation of power 
  gets prominent as the initial deployed power at shorter  scales is increased. For instance, 
  one can observe that when either the value of $n$ in Eq.~(\ref{pw_evol}) is decreased or that of $k_m$ increased 
  for Eq.~(\ref{pw_evol1}) (representing the two choices of 
  initial form of the spectrum)  the evolved spectra shows increasing dominance of the peak.  
We also observe  that the location of the peak depends on the choice of the parameter  $\eta/\tau_m$. In Fig.~\ref{f5} we plot 
the power spectra for  several cases with different values of the parameter $\eta/\tau_m$. It can be seen that there 
is a definite shift of the spectra towards high $k$ values as the value of $\eta/\tau_m$ is reduced. 

To summarize, the main observations of our simulations are (i) a broad non universal character of the evolved spectrum for GHD, 
which shows strong dependence on the initial spectrum (ii) appearance of broad peaks which shift towards longer scales with increasing 
time (iii) with increasing $\eta/\tau_m$ the peak location again shifts towards 
long scales (iv) The peaks are dominant when the initial power content in shorter scales is high. 
In the next section we provide a discussion on these observations.   
\section{Discussion}
\label{diss6}
A hydrodynamic fluid in the incompressible limit is a scale free system as it does not support any normal mode and neither does it 
have any intrinsic length and/or time scale associated with it. 
This results in a power law spectrum at intervening scales where neither forcing nor dissipation is directly effective. 
In contrast, the  incompressible GHD model 
 supports transverse shear wave mode and  
there also exists a special time scale defined by the  relaxation parameter $\tau_m$ in the system. 
The presence of these special scales in the GHD model is responsible for  introducing features different from the  power law 
behavior in the  spectrum.  

 Furthermore, the HD system in 2-D supports two square integral invariants 
in the non dissipative limit, namely energy and enstrophy,  which leads to the characteristic inverse cascade 
phenomena in the context of 2-D HD system.  It is noteworthy here that the 3-D HD system  
does not support the enstrophy invariant and the spectral cascade is direct in this case.  
The GHD system also does not support the enstrophy invariant and hence one would have expected only a direct spectral cascade in this case. 

In the case of GHD, however, we still observe formation of longer scales though it is at a considerable slower time scale  here. 
Furthermore, the structures that form are much shorter than the box size (the observation of peaks in the spectrum testifies to it). 
The accumulation at intervening scales  is 
higher when the initial spectrum has comparatively higher power in short scales. Also, one observes the shifting of peaks towards 
longer scales with time. This suggests that even though the GHD system does not support the second enstrophy 
invariant, there is a 
slow (compared to HD), but perceptible transfer of power towards longer scales. 
We now try to provide a qualitative understanding of this phenomena.  

\begin{figure}[ht]
\centering
\includegraphics[height=10.0cm,width=12.0cm]{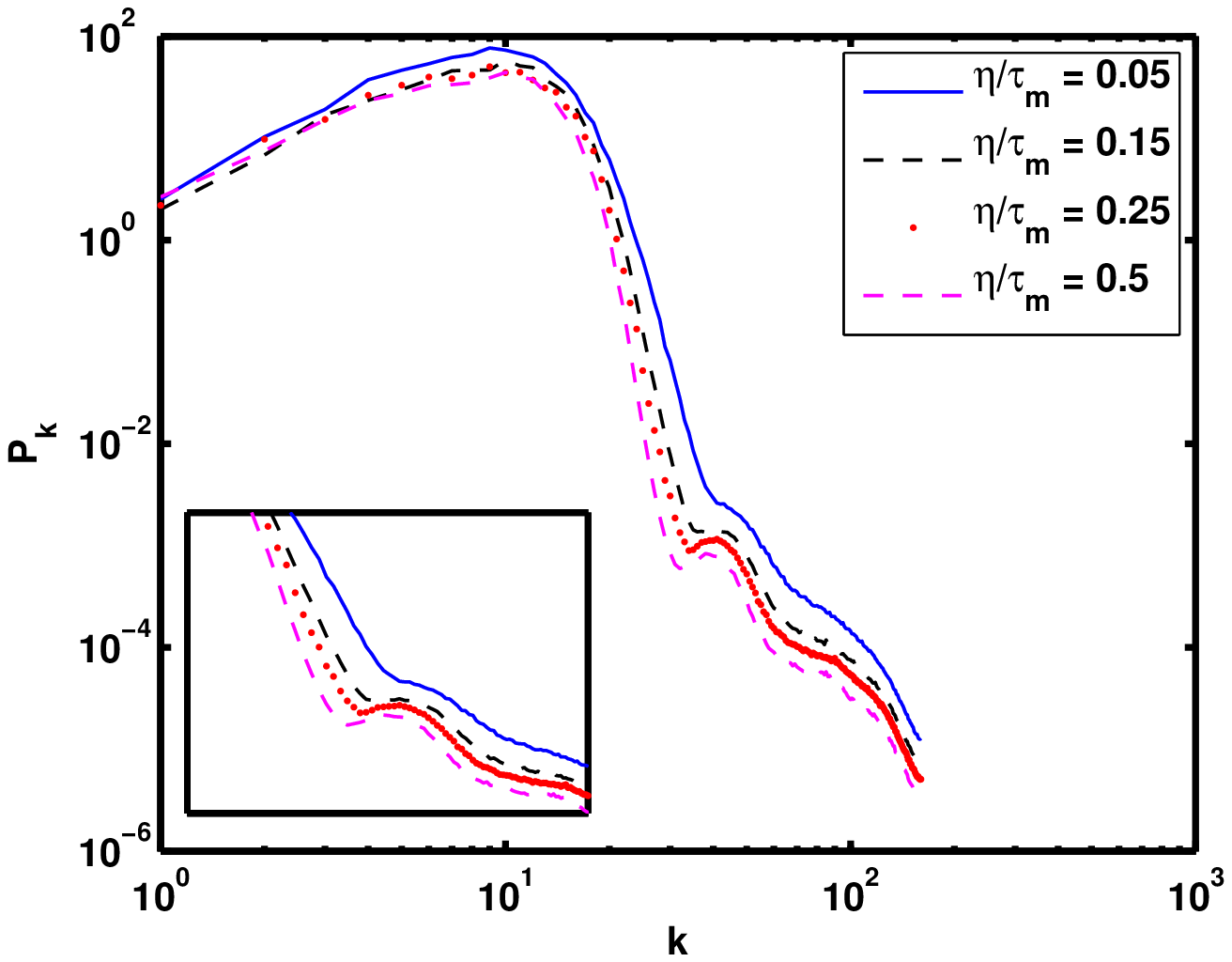}
\caption{Power spectra plots with different value of $\eta/\tau_m$ for incompressible visco-elastic fluid. The initial form of velocity
potential is given as in Eq.~(\ref{pw_evol}) with $n=2$. The spectra for all values of $\eta/\tau_m$ has been plotted here at $t=500$.} 
\label{f5}
\end{figure}

It should be noted that there is an inherent slow dynamics of HD involved in the GHD system of equations, occurring 
at   time scales longer than the memory relaxation time $\tau_m$. This is apparent from Eq.~(\ref{ghd_comp}), when 
the $\tau_m d/dt $ can be ignored in comparison to unity. In this limit the GHD equations reduce to the HD form. 
This suggests that phenomena with typical times scales longer than the memory relaxation time and/or any other time associated 
with the elastic nature of the fluid the GHD system may contribute towards inverse spectral cascade. This is indeed 
what one observes in the simulation.  
There appears, thus, a competition between the HD dynamics of inverse spectral cascade which is 
opposed by the intrinsic elastic behaviour of the system. 
This competition would be characterized in terms of the time scale exceeding those associated with the visco - elastic GHD interactions. 
The GHD is characterized by  the relaxation parameter $\tau_m$.   
{The GHD dynamics also involves the propagation of transverse shear wave with the phase velocity 
$V_p = \sqrt{\eta/\tau_m}$. The time scale intrinsic to the system associated with the transverse shear wave at  any 
given wavenumber $k_s$ would then be 
$t_{sk} \sim 1/k_s V_p$.  Thus at a given time $t$ the 
hydrodynamic response will start playing a role in the GHD system only when the condition $t/t_{sk} >> 1$ gets satisfied. 
An approximate qualitative understanding of the observations can be developed along the following lines. 
For a particular system with predefined $\eta$ and $\tau_m$, $V_p$ is constant. 
Thus at higher wavenumbers $t/t_{sk} >> 1$ has a greater chance to be  satisfied. 
The HD response of perceptible cascade towards longer scales would, therefore, occur from short scale onwards. 
At increasingly longer scales (lower wavenumbers)   the value of  $t_{sk}$ increases making the onset of HD dynamics difficult.
Thus the procedure of  inverse spectral cascade is rendered more and more difficult at longer scales. 
The presence of inverse cascade at short scales and its absence at longer scales  would result in accumulation of power at some scale length. 
However, with increasing time the HD response will slowly keep taking over. 
This results in the continuous slow process of the shifting of the peak (showing power accumulation ) towards longer scales
where the power is observed to accumulate.  

This also explains that as one increases $\sqrt{\eta/\tau_m}$ the hydrodynamic response would set in at decreasing $k$ values. 
Thus, as one increases the value of $\sqrt{\eta/\tau_m}$ the location of the peak shifts towards longer scales as can be seen from Fig.~\ref{f5}.  }

\section{Conclusion}
\label{concl6}
It has been shown that for a visco-elastic medium depicted by the GHD model, turbulence can set in even at low Reynold's number, 
provided the Weisenberg number is high. This is in conformity with experimental observations and other models of elastic turbulence 
studies (e.g. those corresponding to polymeric fluids etc.,).

It is noteworthy that the visco - elastic system considered here supports the  transverse shear wave as a normal mode. 
The question of whether or not the natural modes and scales  of the system have a role  on turbulence has continued to remain an outstanding problem. 
Attempts have so far been made on the basis of identifying the differences  predicted theoretically on the power spectral index of a stationary 
turbulence state. This has so far proved an extremely  difficult exercise and has resulted in endless controversies without settling 
the issue one way or the other. 

Against this backdrop we have shown that for the case of turbulence in 
visco - elastic medium governed by the Generalized Hydrodynamic equation,  
the  quasi - stationary behavior of the turbulent spectrum provides ample evidence of  
the involvement of the memory relaxation time $\tau_m$ and the transverse shear wave in the spectral cascade process. 

Furthermore, the appearance of shorter scales in 2-D for GHD system is also 
suggestive of enhanced transport and mixing properties of the GHD system vis. a vis 
that of 2-D  HD dynamics.


\bibliographystyle{unsrt} 
\bibliography{sanat_ghd_chap6}

\end{document}